\documentstyle[preprint,eqsecnum,aps,epsf]{revtex}
\newif\iftightenlines\tightenlinesfalse
\tightenlines\tightenlinestrue
\begin{document}
%
\def\tG{\tilde G}
\def\ETC{E_T^c}
\def\eslt{\not\!\!{E_T}}
\def\to{\rightarrow}
\def\te{\tilde e}
\def\tf{\tilde f}
\def \tlam{\tilde{\lambda}}
\def\tl{\tilde l}
\def\tb{\tilde b}
\def\tst{\tilde t}
\def\tt{\tilde t}
\def\ttau{\tilde \tau}
\def\tmu{\tilde \mu}
\def\tg{\tilde g}
\def\tga{\tilde \gamma}
\def\tnu{\tilde\nu}
\def\tell{\tilde\ell}
\def\tq{\tilde q}
\def\tw{\widetilde W}
\def\tz{\widetilde Z}
\def\Rsl{\not\!\!{R}}
%
%
\preprint{\vbox{\baselineskip=14pt%
   \rightline{FSU-HEP-000328}\break 
   \rightline{UH-511-958-00}
}}
\title{THE REACH OF THE CERN LARGE HADRON COLLIDER
FOR GAUGE-MEDIATED 
SUPERSYMMETRY BREAKING MODELS}

\author{Howard Baer$^1$, P.~G. Mercadante$^1$,
Xerxes Tata$^2$ and Yili Wang$^2$}
\address{
$^1$Department of Physics,
Florida State University,
Tallahassee, FL 32306, USA
}
\address{
$^2$Department of Physics and Astronomy,
University of Hawaii,
Honolulu, HI 96822, USA
}
%
\maketitle
\begin{abstract}

We examine signals for sparticle production at the CERN Large Hadron 
Collider (LHC) within the
framework of gauge mediated supersymmetry breaking models with a low
SUSY breaking scale for four
different model lines, each of which leads to qualitatively different
signatures. We first examine the reach of the LHC via the canonical
$\eslt$ and multilepton channels that have been advocated within the
mSUGRA framework. Next, we examine special features of each of these
model lines that could serve to further enhance the SUSY signal over
Standard Model backgrounds. We use ISAJET to evaluate the SUSY reach of
experiments at the LHC. We find that the SUSY reach, measured in terms
of $m_{\tg}$, is at least as large, and sometimes larger, than in the
mSUGRA framework. In the best case of the co-NLSP scenario, the reach
extends to $m_{\tg} \geq 3$~TeV, assuming 10 $fb^{-1}$ of 
integrated luminosity.

\end{abstract}

\medskip



\section{Introduction}

The search for supersymmetric particles has become a standard item for
on-going as well as future experiments at high energy colliders. No
signal has as yet been found, and the absence of evidence has been
translated to lower limits on the masses of various
sparticles~\cite{exprev}.  Since the qualitative features as well as the
size of experimental signals are determined by sparticle production
cross sections and decay patterns, any such interpretation can strictly
speaking only be done within the framework of a particular model, most
often taken to be the minimal supergravity (mSUGRA) model. While
it is true that in some cases the non-observation of a signal can be
translated into a sparticle mass bound in a relatively model-independent
manner\footnote{For example, at LEP, it is possible to infer a bound
on the mass of the right handed slepton, assuming only that it is the
lightest unstable sparticle, so that it must decay via $\tell_R \to \ell
\tz_1$; even here it is tacitly assumed that the decay of $\tell_R$ is
prompt.}, the extraction of signals (or bounds) becomes increasingly
complicated if several sparticles are expected to be simultaneously
produced as, for instance, at hadron colliders such as the CERN LHC. 
This may be due to significant
differences of sparticle mass and decay patterns from  the mSUGRA
expectations~\cite{difference}, or due to new decay modes being
accessible as in gauge-mediated SUSY breaking (GMSB) models~\cite{gmsb}
with a low SUSY breaking scale, so that the gravitino is
by far the lightest supersymmetric particle (LSP).

In recent years, there has been a flurry~\cite{flurry,bmtw,tevatron,frank2,mt}
of theoretical as well as experimental activity~\cite{tevgmsb,lepgmsb}
on the detection of SUSY signals within the GMSB framework, where SUSY
breaking is communicated from the hidden sector to the observable sector
via Standard Model (SM) gauge interactions of messenger particles
(which, we assume, can be classified into $n_5$ complete vector
representations of $SU(5)$) with quantum numbers of $SU(2)$ doublets of
quarks and leptons whose mass scale is characterized by $M$. The soft
supersymmetry breaking masses for the SUSY partners of SM particles are
thus proportional to the strength of their gauge interactions, so that
squarks are heavier than sleptons, while the gaugino masses
satisfy the usual ``grand unification'' mass relations,
though for very different reasons. Within the minimal version of this
framework, the couplings and masses of the sparticles in the observable
sector are determined (at the messenger scale $M$) by the parameter set,
\begin{equation}
\Lambda,M,n_5,\tan\beta,sign (\mu ),C_{grav}.
\label{parset}
\end{equation}
The parameter $\Lambda$ sets the scale of sparticle masses and is the
most important of these parameters. The model predictions for soft-SUSY
breaking parameters at the scale $M$ are evolved to the weak
scale. The parameter $C_{grav} \geq 1$ \cite{bmtw}
and enters only into the partial
width for sparticle decays to gravitino. The decay to a gravitino is
phenomenologically important only for the decay of the next to lightest
supersymmetric particle (NLSP). We fix $C_{grav}=1$ which corresponds to
the fastest possible decay of the NLSP into the gravitino: for larger
values of $C_{grav}$, the NLSP may decay with an observable decay gap
(possibly improving the sensitivity to the signal),
or may even pass all the way through the detector. 

The phenomenological implications of the model are only weakly sensitive
to $M$. Aside from the scale of sparticle masses, the single item that
most sensitively determines sparticle signals at colliders is the
identity of the NLSP. In the simplest version of model, the NLSP is
either the lightest neutralino ($\tz_1$) with the hypercharge gaugino as
its dominant component, or the lighter stau ($\ttau_1$). In the latter
case, depending on the value of $\tan\beta$, $\te_1$ and $\tmu_1$ may be
essentially degenerate with $\ttau_1$. 

If the NLSP is a $U(1)$ gaugino-like neutralino, it dominantly decays
via $\tz_1 \to \gamma \tG$, and SUSY events always contain two hard,
isolated photons and $\eslt$ (from the escaping gravitinos) in addition
to jets and/or leptons. In the case where the NLSP is a stau which is
significantly lighter than other sleptons, it is produced in most SUSY
decay chains, and SUSY events in this scenario typically contain several
tau leptons. Other light sleptons decay via $\tell_1 \to \ell \tz_1$ if
the decay is allowed; otherwise as discussed in Ref. \cite{tevatron},
these mostly decay via $\tell_1^- \to \ttau_1^+\ell^-\tau^-$ or
$\tell_1^- \to \ttau_1^-\ell^-\tau^+$ which (even for $C_{grav}=1$), as
long as these are not kinematically strongly suppressed, have a much
larger branching ratio than for the decay $\tell_1 \to \ell \tG$.  For
the case where $m_{\tell_1}-m_{\ttau_1} \leq m_{\tau}$ (the co-NLSP
scenario), only the gravitino mode is allowed, so that we may then
expect a significant multiplicity of isolated $e$'s and $\mu$'s in SUSY
events.

In a previous paper \cite{tevatron}, we examined SUSY signals and the
reach of the Fermilab Tevatron Main Injector (for integrated
luminosities of 2 and 25~$fb^{-1}$) for representative cases for each
one of these three model lines discussed in the previous paragraph. We
also examined the reach for the so-called higgsino model
line~\footnote{These four model lines were first proposed for study at
the Run II SUSY and Higgs Workshop held at Fermilab in 1998. The
nomenclature used is different from that in the Workshop.} where the
NLSP is a higgsino-like neutralino. Although this does not occur in the
minimal version of the GMSB framework because $\mu$ tends to be large
compared to the hypercharge gaugino mass $M_1$, we were motivated to
examine this case since experimental signals tend to be quite different
because the NLSP dominantly decays via $\tz_1 \to h \tG$ (rather than
$\tz_1 \to \gamma \tG$) unless this decay is suppressed by phase
space. Our purpose here is to perform a similar study for the reach of
the CERN Large Hadron Collider (LHC), a $pp$ collider with a center of mass
energy $\sqrt{s}=14$~TeV, currently scheduled to begin operation in
2005. We assume throughout this paper that an integrated luminosity 
of 10 fb$^{-1}$ is collected, which corresponds to a year of running at
the lower value of the design luminosity.

Toward this end we examine the SUSY reach of the LHC for,
\begin{itemize}
\item model line A: $n_5=1$, $M=1000$~TeV, $\tan\beta = 2$, $\mu > 0$,
which gives a $U(1)$ gaugino-like NLSP;
\item model line B: $n_5=2$, $M=3\Lambda$, $\tan\beta=15$, $\mu > 0$,
for which the NLSP is a stau;
\item model line C: $n_5=3$, $M =3\Lambda$, $\tan\beta=3$, $\mu>0$,
which yields the co-NLSP scenario;
\item model line D: $n_5=2$, $M=3\Lambda$, $\tan\beta=3$,
$\mu=-\frac{3}{4}M_1$, which gives a higgsino-like neutralino as the
NLSP, which if heavy enough mainly decays via $\tz_1 \to h\tG$;
\item model line E: $n_5=2$, $M=3\Lambda$, $\tan\beta=3$,
$\mu=\frac{3}{4}M_1$, which also gives a higgsino-like neutralino as the
NLSP, but which mainly decays via $\tz_1 \to Z\tG$ as long as the decay
is not suppressed by phase space.
\end{itemize}
Since we have already examined the LHC reach for model line A
before~\cite{bmtw}, we will confine our attention to model lines B-E in
this paper. The Tevatron reach for model lines B, C and D was studied in
Ref. \cite{tevatron}. The reach for a gaugino-like NLSP model line, but
with a slightly different set of parameters (which should have no
qualitative effect on the results) was also examined in this same
study. Model line E is new. The difference between D and E is that while
$\tz_1$ mainly decays via $\tz_1 \to h\tz_1$ for model line D, the decay
$\tz_1 \to Z\tG$ dominates~\cite{mt} in the case of model line E
(assuming, of course, that these decays are kinematically unsuppressed).

For each of the model lines B-E, we first examine the reach using the
strategy~\cite{lhc1,lhc2} for SUSY searches within the mSUGRA framework:
{\it i.e.} the reach via the $\eslt$ plus $0\ell$, $1\ell,2\ell$ and
$3\ell$ inclusive channels. We expect that this (or something like it)
will be the canonical strategy~\cite{CMS,ATLAS} for SUSY searches in the
initial data samples from the LHC. For each of these model lines, we
then examine whether and by how much the reach can be improved using the
special features characteristic of the particular model line, {\it e.g.}
the presence of hard, isolated photons in model line
A~\cite{bmtw}.

The remainder of this paper is organized as follows. In Sec.~II, we
briefly discuss our simulation using the program
ISAJET~\cite{isajet}. In Section III, we discuss the LHC reach for the
four model lines B-E. We end with a summary and some concluding remarks
in Sec.~IV.

\section{ Event Simulation }

We use the program ISAJET v 7.43~\cite{isajet} for our simulation of
SUSY events. This incorporates decay matrix elements for three body
decays of gluinos, charginos and neutralinos in the event generator.
The implementation of the GMSB framework into ISAJET v 7.43 has been
described elsewhere~\cite{tevatron,isajet} and will not be repeated
here. ISAJET uses the GMSB model predictions for sparticle masses
as initial conditions valid at the messenger
scale $M$, then evolves these down to the weak scale relevant for
phenomenology, and finally calculates the `MSSM parameters' that are
then used in the evaluation of sparticle cross sections and decay
widths.

For detector simulation at the LHC, we use the ISAJET toy calorimeter
simulation package CALSIM. We simulate calorimetry covering $-5 \leq
\eta \leq 5$ with a cell size given by $\Delta\eta \times \Delta\phi=
0.05 \times 0.05$, and take the hadronic calorimeter
resolution to be $50\% /\sqrt{E}\oplus 0.03$ for $|\eta |<3$, where
$\oplus$ denotes addition in quadrature, and to be $100\%
/\sqrt{E}\oplus 0.07$ for $3<|\eta |<5$, to model the effective $p_T$
resolution of the forward calorimeter including the effects of shower
spreading, which is otherwise neglected. We take electromagnetic
resolution to be $10\% /\sqrt{E}\oplus 0.01$.  Although we have included
these resolutions, which are typical of ATLAS~\cite{ATLAS} and
CMS~\cite{CMS}, we have made no attempt to estimate the effects of
cracks, edges, and other problem regions. Much more detailed detector
simulations are needed to understand the effects of such regions and of
the resulting non-Gaussian tails, particularly on the $\eslt$
resolution.

Jets are defined as hadronic clusters with $E_T>25$~GeV within a cone of
$\Delta R= \sqrt{\Delta\eta^2+\Delta\phi^2} = 0.7$ with $|\eta_j| \leq
3.5$. Muons and electrons with $E_T > 10$~GeV and $|\eta_{\ell}| < 2.5$
are considered to be isolated if the the scalar sum of electromagnetic
and hadronic $E_T$ (not including the lepton, of course) in a cone with
$\Delta R=0.3$ about the lepton to be smaller than 5 GeV.  Isolated
leptons are also required to be separated from one another by $\Delta R
\geq 0.3$.  Photons are identified as isolated with $|\eta_{\gamma}|<
2.5$ and $E_T > 10$~GeV if the additional $E_T$ within a cone of
$\Delta R = 0.3$ about the photon is less than 2.5~GeV. For
$b$-jets, we require a jet (satisfying the above jet criteria) to
contain a $B$-hadron with $p_T \geq 15$~GeV and $|\eta_j| \leq 3$.  Jets
with $E_T \geq 50$ GeV are tagged~\footnote{For the purposes of
reconstructing $h$ produced via $\tz_1 \to h\tG$ for model line D, we
assume that softer $b$-jets can also be tagged, but with a reduced
efficiency as detailed in Sec.~III C.} as a $b$-jet with a
probability of $50\%$. QCD jets with $E_T = 50$ ($\geq 100$) GeV are
misidentified as $b$-jet with a probability of $0.79\%$ ($2\%$) with
a linear interpolation in between~\cite{ratio}.

\section{The SUSY Reach of the LHC}

In this section we identify strategies to enhance the LHC SUSY reach for
GMSB models beyond what would be obtained via canonical multijet +
multileptons + $\eslt$ analyses \cite{lhc1,lhc2,ratio} advocated for the
mSUGRA framework. We examine the reach for model lines B-E, for which
the nature of the NLSP, and hence the characteristic SUSY signals, vary
with the model line. The reach of model line A has already examined
previously.~\cite{bmtw}. We use ISAJET v 7.43 to
compute signal cross sections, after incorporating acceptances  to
simulate the experimental conditions at the LHC together with
additional cuts that serve to separate the SUSY signal from SM
backgrounds.

\subsection{Model Line B: The Stau NLSP model line}

To realize the case of the stau NLSP, we take $n_5=2$, $M=3\Lambda$,
$\mu > 0$, and choose $\tan\beta=15$ to ensure that the stau is light
enough for other sleptons to be able to decay to it. It is worth noting
that although $m_{\ttau_1} < m_{\tell_1}$, the sleptons are all
sufficiently close in mass so that generally speaking the phase space
for chargino and neutralino decays to staus and other sleptons is not very
different. For $\Lambda \agt 30$ TeV, $m_{\ttau_1} \alt m_{\tz_1}$, but
for $\Lambda \alt 31$ TeV, $\tz_1\to \tau\ttau_1$ is kinematically
forbidden, and $\tz_1$ would decay via the four body decay $\tz_1 \to
\nu_{\tau}\ttau_1 W^*$ (which is not yet included in ISAJET) or via
$\tz_1 \to \gamma\tG$. In our study, we consider $\Lambda \geq 35$ TeV,
the range safe from LEP constraints.

In Fig. \ref{csectionB}, we show cross sections for various sparticle
production processes at the LHC. For $\Lambda \leq 125$~TeV ($m_{\tg}
\alt 1.8$~TeV), squark and gluino production (solid line) dominates,
while for larger values of $\Lambda$ electroweak production of various
charginos and neutralinos (dashed line) are the dominant SUSY process.
For cross sections in the
range of interest at the LHC, direct slepton and sneutrino production
occurs at $\sim \frac{1}{3}$ the rate for chargino and neutralino
production.  Cascade decays of squarks and gluinos naturally lead to
events with $n$ leptons + $m$ jets + $\eslt$. These event topologies may
also arise from the production of gluinos and squarks in association
with a chargino or a neutralino, and also from chargino and neutralino
production.

The decay patterns of charginos, neutralinos and sleptons have been
discussed in detail in Ref. \cite{tevatron} and we will only summarize
the results here.  The two body decay $\tw_1 \to \ttau_1 \nu$ is always
accessible, and dominates for $\Lambda \leq 45$~TeV ($m_{\tw_1} \alt
210$~GeV). The decay rate for $\tw_1 \to W\tz_1$ becomes comparable to
that for $\tw_1 \to \ttau_1 \nu$ for $\Lambda \sim 60$~TeV, while for
very heavy charginos ($m_{\tw_1} \agt 900$~GeV), the branching ratio for
$\tw_1 \to W\tz_1$ is around 75\%.  Turning to neutralinos, $\tz_2 \to
\ttau_1 \tau$ is the main decay mode of $\tz_2$ if $m_{\tz_2} \alt
300$~GeV. The decay $\tz_2 \to \tz_1 h$ is only important for $\Lambda
\agt 55$~TeV ( $m_{\tz_2} \agt 270$~GeV) and becomes dominant for
$\Lambda \agt 80$~TeV ( $m_{\tz_2} \agt 400$~GeV).  Branching fractions
for $\tz_2\to \tell_1\ell$ ($\ell=e,\mu$) may also be significant and vary
from 20\% each for $\Lambda = 35$~TeV to $< 1$\% for $\Lambda
=180$~TeV. The branching fraction for the decay $\tz_2 \to Z\tz_1$ is
$\sim 5$\% if this decay is kinematically accessible.  For small enough
$\Lambda$, the lightest neutralino $\tz_1$ mainly decays via $\tz_1 \to
\ttau_1\tau$, but for large enough values of $\Lambda$ its decay to
other sleptons are also important, and together exceed the decays to
stau when for $\Lambda > 80$~TeV ($m_{\tz_1}\alt 220$~GeV).  Finally,
for small values of $\Lambda$ accessible at Tevatron upgrades, $\tell_1$
can only decay via $\tell_1 \to \ell \tz_1$. For larger $\Lambda$, this
channel is closed, the neutralino is virtual and then the decays
$\tell_1^- \to \ttau_1^+ \ell^-\tau^-$ and $\tell_1^- \to \ttau_1^-
\ell^-\tau^+$ dominate. However, the decay $\tell_1 \to \ell\tG$ also becomes
significant (branching fraction $\geq$10\% for $\Lambda \agt 120$~TeV)
if $\Lambda$ is large enough.

We begin our discussion of the LHC reach via canonical searches in the
multijet plus multilepton plus $\eslt$ channels as discussed in
Ref. \cite{lhc1,lhc2}. For squark and gluino production, the $p_T$ of
the primary jets from gluinos (squarks decay to gluinos), as well as the
$\eslt$, are expected to scale with $m_{\tg}$. Since the signal
decreases with increasing $m_{\tg}$, its separation from SM backgrounds
can be optimized by requiring cuts on jets and $\eslt$ that scale with
the gluino mass.  The momenta of leptons, produced far down in the
cascade decay chain from chargino and neutralino daughters, do not scale
in energy the same way as $\eslt$ or energies of jets which are produced in the
first step of the decay cascade.  Following Ref.~\cite{lhc1,lhc2}, we
introduce the running cut variable $E_T^c$, and require that the signal 
events satisfy:
\begin{itemize}
\item $n_{jet} \geq 2$ (with $E_{T(j)}> 100$~GeV);
\item the transverse sphericity, $S_T \ge 0.2$;
\item $E_T(j_1)$, $E_T(j_2) >E^c_T$ and $\eslt > E^c_T$;
\item for $1 \ell$ signal, $p_T(\ell)> 20$ GeV and $M_T(\ell, \eslt)>
100$~ GeV; 
\item  for $n \geq 2$ lepton signal, $p_T(\ell_1, \ell_2)> 20$~GeV,
\end{itemize}
in addition to the basic acceptance cuts.  We classify signal events
according to their lepton topology into the nonleptonic $\eslt$ + jets
channel, the opposite-sign (OS) dilepton + jets +$\eslt$ channel, the
same-sign (SS) dilepton + jets +$\eslt$ channel, and the multilepton +
jets +$\eslt$ channel with $n_\ell \geq 3$.  Throughout this paper, we
consider a signal to be observable in a particular channel if there are
at least 5 signal events and also the signal has a statistical
significance $\geq 5$, assuming 10 fb$^{-1}$ of integrated luminosity. 
We study the observability by varying $E_T^c$
between 100 and 500~GeV in steps of 100~GeV. 

The dominant SM backgrounds are $W$, $\gamma^*$ or $Z$ + jet production,
$t\bar{t}$ production and vector boson pair production. Since our
analysis cuts are identical to those in Ref.~\cite{lhc1,lhc2}, we do not
re-compute these backgrounds but instead use the background cross
sections as a function of $E_T^c$ given in these papers in our
analysis.~\footnote{We caution the reader that these backgrounds were
computed using CTEQ2L structure functions, while for the signal 
CTEQ3L\cite{cteq} structure functions have been used.}

The results of our computation of the reach of the LHC using the mSUGRA
cuts are summarized in Fig~\ref{reachmlB} where we show the maximum
$\sigma_S/\sqrt{\sigma_B}$ as a function of $E_T^c$, requiring further
that there be at least 5 signal events per 10~$fb^{-1}$.  For each value
of $\Lambda$ that we studied, the number on the various lines denotes
the $E^c_T$ value for which the signal is observable according to our
criteria, and for which the signal significance is the greatest (subject
to the five event minimum), i.e. a 2 denotes $E^c_T=200$~GeV yields the
best significance, {\it etc.}  In other words, this number provides
information about the optimal strategy for the particular value of
$\Lambda$.  The different lines correspond to the different event
topologies: $0 \ell + \eslt$ (lower solid), $1\ell+\eslt$ (upper solid),
SS dileptons + $\eslt$ (dashed), OS dileptons + $\eslt$ (dotted) and
trileptons (dot-dashed).  The horizontal line denotes the $5 \sigma$
level for an integrated luminosity of 10~$fb^{-1}$.

We see from the figure that unlike the case of the mSUGRA framework, the
inclusive $3\ell$ channel provides the greatest reach. This is
presumably because of additional leptons that come from slepton decays
to the gravitino, or as secondaries of daughter taus produced via the
decay of the stau NLSP. The $\eslt$ signal (with a lepton veto) gives a
rather poor reach for essentially the same reason.  We can probe
$\Lambda$ values beyond 150 GeV which corresponds to a gluino in excess
of 2 TeV in this channel. Notice also that if $\Lambda \leq 125$~TeV
($m_{\tg} \alt 1.8$~TeV) confirmatory signals should appear in several
other channels. For comparison, we note that even for an integrated
luminosity of 25~$fb^{-1}$, the reach of the Tevatron was
projected~\cite{tevatron} to be about 50~TeV.

Next, we examine whether we can extend the reach by requiring the
presence of additional tagged $\tau$ leptons.  In Table I, we show the
signal cross section for $\Lambda = 120$ and 160~TeV, with events
classified first by the number of identified taus, and then by the
lepton multiplicity. Here, a jet with $E_T \geq 40$~GeV is identified as
a tau if it has one or three charged prongs with $p_T\geq 2$~GeV in a
10$^{\circ}$ cone about the jet axis, with no other tracks in the corresponding
30$^{\circ}$ cone. If the number of charged prongs is three, we require their
net charge to be $\pm 1$ and the invariant mass to be smaller than
$m_{\tau}$. QCD fakes are included with the same algorithm used in
Ref. \cite{tevatron}. The cross sections shown in Table~I are obtained
with just $\eslt \geq 50$~GeV in addition to the basic lepton and $\tau$
identification cuts. We see that even with this minimal requirement, the
cross section for $n_e+n_{\mu}+n_{\tau} \geq 4$ events is just 3.25
(.92)~$fb$ for $\Lambda = 120 (160)$~TeV. Since these channels all suffer from
physics, and more importantly instrumental, backgrounds from SM sources,
we conclude that tagging hadronically decaying taus is unlikely to
improve the reach for this model line: cuts to remove these will reduce
the signal to below the observable level for $\Lambda$ values where the
canonical strategy does not lead to a detectable signal.

We also attempted to extract the signal via ``clean'' multilepton plus
multi-tau channels where we veto events with jets that are not
identified as taus. We found that typically the signal reduces by an
order of magnitude (or more, depending on the transverse momentum
threshold for the jet veto) and falls well below the one event level in
all the channels. We thus conclude that the canonical SUSY search strategy
\cite{lhc2} via multilepton channels provides the largest reach for
model line B.

\subsection{Model Line C: The co-NLSP Model Line.}

For this model line, $\te_1$, $\tmu_1$ and $\ttau_1$ are approximately
degenerate, and $\te_1$ and $\tmu_1$ cannot decay to $\ttau_1$.  To
realize this, we choose $n_5=3$ and $\tan\beta=3$ with other parameters
as before. Various sparticle production cross sections are shown in
Fig.~\ref{csectionC}. We see that while production of squarks and
gluinos dominates for $\Lambda \leq 125$~TeV ($m_{\tq} \leq 2.3$~TeV,
$m_{\tg} \leq 2.5$~TeV), slepton pair production is dominant
for $\Lambda \agt 125$~TeV ( corresponding to $m_{\tl} \agt 380$~GeV).

In this model line, gluinos decay directly to squarks. 
The left-squarks typically cascade to $\tw_1$
and $\tz_2$ which then decay to the lightest neutralino, while $\tq_R$
mostly decays directly to $\tz_1$.  The sparticle decay patterns have
been discussed in Ref.~\cite{tevatron}.  The lightest neutralino decays
via $\tz_1 \to \tell_1 \ell$, ($\ell= e$, $\mu$, and $\tau$), with
branching fractions essentially independent of the lepton
flavour. Sleptons and sneutrinos decay via $\tf_2 \to f\tz_1$ while
$\tf_1 \to f\tG$ ($f =\ell$,$\nu$). For very small $\Lambda$ (excluded
by LEP 2 since the slepton is then very light), the charginos can only
decay via $\tw_1 \to \ttau_1\nu_{\tau}$~\footnote{This decay proceeds
mainly via the higgsino component; as a result, decays to other slepton
flavours are strongly suppressed.} but over much of the allowed range of
$\Lambda$ the decays $\tw_1 \to \tnu\ell,\nu\tell$ dominate and have
(approximately) flavour-independent branching fractions. The decay
$\tw_1\to W\tz_1$ also becomes significant for charginos heavier than
$\agt 200$~GeV, and has a branching fraction that decreases from a
maximum of $\sim
30$\% to $\sim 15$\% as $\Lambda$ increases to 130~TeV.  The second
lightest neutralino $\tz_2$ decays to sleptons with branching fractions
more or less independent of the lepton flavour. Decays to the heavier
(mainly left-handed) sleptons and sneutrinos dominate when these are
not kinematically suppressed. The branching fraction for the decay
$\tz_2 \to \tz_1 h$ is also significant once this decay is allowed, and
the branching fraction reduces from a maximum of 25\% to below 15\% for
$\Lambda =130$~TeV. Neutralino decays to $Z$ are negligible. An
inspection of these decay patterns shows that it would be natural to
expect events with a large multiplicity of isolated
leptons( $e$ and $\mu$) from sparticle production, so that multilepton
plus $\eslt$ events should provide the best reach. 

The naturally high lepton multiplicity in these events led us to focus
on the essentially background-free $n_{\ell} \geq 4$ lepton plus $\eslt$
channel. We have not attempted to make a quantitative calculation of the
background but have instead contented ourselves with trying to bound
cross sections for physics sources of such events. 
These include, 
\begin{itemize}
\item $4t$ production with $\sigma(4t)=6$~$fb$\cite{bsp}, which yields
$\sigma(4\ell) = 0.014$~$fb$ (neglecting isolated leptons~\footnote{The
cross section for 3 leptons from the decays of $W$s is 0.25~$fb$; since
the branching fraction for leptonic $b$ decays is about 0.22, we
conclude that as long as the lepton from $b$ is isolated less than $\sim
1/20$ of the time, this background is smaller than the $4\ell$ level
estimated above.} from $b$ decays);
\item $4W$ production with $\sigma(4W) = 1$~$fb$~\cite{bhp}, which
yields $\sigma(4\ell) = 2.3 \times 10^{-3}$~$fb$;
\item $WWZ$ production with $\sigma(WWZ)=150$~$fb$~\cite{bh}, which
yields $\sigma(4\ell)=2.4\times 10^{-2}$~$fb$ (here we assume that
events $Z \to \ell\bar{\ell}$ can be removed by a mass cut, and retain
leptons only from $Z \to \tau\bar{\tau}, \tau \to \ell\nu\bar{\nu}$);
\item $ZZ$ production with $\sigma(ZZ)=10$~pb, which yields
$\sigma(4\ell)= 0.11$~$fb$ where, once again, we assume that the leptons
come only via $Z\to \tau\bar{\tau}$ decays. These events tend not to have
a large $\eslt$ and softer leptons, and the background can be greatly
reduced by a cut on $\eslt$ and the lepton $p_T$.
\end{itemize}

Turning to the signal, we show the distributions of the two hardest
leptons in $\geq 4\ell$ SUSY events in Fig.~\ref{mlC_lep12pt} for four
choices of $\Lambda$. We see that requiring these leptons to be harder
than 50~GeV reduces the signal by just a few percent for
$\Lambda=120$~TeV. Even for $\Lambda=35$~TeV, the reduction is just
40\%, which is very affordable because of enormous SUSY cross sections
at the LHC. Although we do not show this, we have also checked that the
signal loss from requiring $\eslt \geq 100$~GeV is again only a few
percent for $\Lambda=120$~TeV.

To obtain the LHC reach via the $\geq 4\ell$ signal for model line C, we thus
require:
\begin{itemize}
\item $\eslt > 100$~GeV; 
\item $p_T(\ell_1)$, $p_T(\ell_2)> 50$~GeV;
\item a $Z$ veto -- we remove events where with $\ell^+\ell^-$ pairs
that reconstruct the $Z$ mass to within 10~GeV. 
\end{itemize}
We assume that these cuts reduce the SM physics backgrounds already
found to be below 0.2~$fb$ to completely negligible levels. Assuming
that instrumental backgrounds are under control, we then estimate the
reach for this model line by requiring at least five signal events in
10~$fb^{-1}$ as illustrated in Fig.~\ref{reachmlC} where we show the
signal cross section with $n_\ell \geq 4$ versus $\Lambda$.  We see that
the reach extends up to $\Lambda = 155$~TeV, which corresponds to
squarks (gluinos) just below (above) 3~TeV, to be compared with the
reach of $\Lambda \sim 55$~TeV \cite{tevatron} for an integrated
luminosity of 25~$fb^{-1}$ at the Tevatron.

Since the sleptons are the co-NLSPs in this scenario, we have also
examined the range of parameters where these can be directly observed at
the LHC. We follow the general strategy for slepton searches at the LHC
detailed in Ref.~\cite{slp}, with some modifications to optimize for the GMSB
scenario. The analysis cuts that we impose are,
\begin{itemize}
\item exactly two isolated same flavor OS leptons, each with $p_T(l) \ge
20$~GeV;
\item $\eslt> 100$~GeV;
\item number of jets with $E_{Tj} \geq 25$~GeV is zero (jet veto);
\item $\Delta \phi(p_{T(\ell\bar \ell^`)}, \eslt) \geq 160^\circ$.
\end{itemize}
These cuts above are exactly the same as in Ref.~\cite{slp}. For the final
cut, however, we choose instead,
\begin{itemize}
\item $p_T(\ell_1)> 80$~GeV, $p_T(\ell_2)> 60$~GeV and $\Delta\phi(\ell_1,
\ell_2)< 140^{\circ}$.
\end{itemize}

The major SM background at the LHC come from $t\bar{t}$ production and $WW$ 
production. The jet veto is crucial to reduce the top quark
background. In our simulation,
we have assumed that jets can be vetoed with an efficiency of 99\%. After these
cuts, the $t\bar{t}$ background is 0.06
$fb$ , while the $WW$ background is 0.038 $fb$, so that the total SM
background $ \sim 0.1~fb$ and the minimum detectable cross section
corresponds to 5 events per 10~$fb^{-1}$.

The signal cross section for opposite sign dilepton from just slepton
production (slepton, chargino and neutralino production) is shown by the
dashed (solid) line in Fig.~\ref{reachslepC}. We see that except for the
$\Lambda=30$~TeV case (corresponding to $m_{\tell_1}=99$~GeV, just
beyond the current bound from LEP 2) the signal dominantly comes from
slepton pair production because $m_{\tw_1} \sim m_{\tz_2} \gg
m_{\tell}$.  The reason the dashed curve falls for low values of
$\Lambda$ is the inefficiency of our cuts (which have been optimized for
heavier sleptons) for such light sleptons. Nonetheless, even with these
cuts~\footnote{Other cuts detailed in Ref.~\cite{slp} would be better
suited for detection of these lighter sleptons, and also for separating
the slepton signal from the signal from charginos.} there is ample
signal for SUSY not to evade detection at the LHC.  We see that the
reach extends up to $\Lambda = 90$~TeV corresponding to $m_{\tell_1}
\simeq 280$~GeV. The reason that the reach in slepton mass is
significantly smaller than the corresponding reach within the mSUGRA
framework is because the production of $\tell_L$ and $\tnu_L$, which was
the main contributor to the slepton cross section in mSUGRA, is now
kinematically suppressed: {\it i.e.} within the GMSB, $\tell_R\tell_R$
production forms the bulk of the slepton cross section.

\subsection{Model Line D: A Higgsino NLSP with $\tz_1\to h\tG$}

Within the minimal GMSB framework, the lightest neutralino is dominantly
the hypercharge gaugino. However, since the phenomenology is very
sensitive to the nature of the NLSP, we entertain the possibility that
in other models, the NLSP may be higgsino-like. We do not attempt to
construct any model, but take a purely phenomenological approach, by
letting $\mu$ to be a free parameter instead of fixing it via radiative
electroweak symmetry breaking conditions.~\footnote{For instance,
additional interactions needed to generate $\mu$ and also the
$B$-parameter in this framework could conceivably alter the relation
between $\mu$ and the gaugino masses.} If $\mu$ is smaller than the
hypercharge gaugino mass $M_1$, the NLSP will indeed contain a large
higgsino component. We take $n_5=2$, $\tan\beta=3$, $M/\Lambda =3$,
$C_{grav}=1$ but fix $\mu =-\frac{3}{4}M_1$ rather than the value
obtained from radiative electroweak symmetry breaking.  Within this
`small $\mu$' model line, the two lightest neutralinos and the lighter
chargino are Higgsino-like and close in mass, while the heavier
charginos and neutralinos are gaugino-like. The $\tz_1$ is the NLSP and
the $\tz_2$ and $\tw_1$ are only little heavier. The fermions from
$\tw_1$ and $\tz_2$ decays to $\tz_1$ will be rather soft. For this sign
of $\mu$, the NLSP dominantly decays via $\tz_1 \to h\tG$ if the $\tz_1$ 
is heavy
enough ($\Lambda \agt 80$~TeV) so that the decay is not suppressed by
phase space. The mass $m_h$ of the lighter Higgs
boson is just above 100~GeV, independent of $\Lambda$.

If $\Lambda \agt 50$~TeV (smaller values of $\Lambda$ would have
charginos and neutralinos accessible at LEP 2), we see from
Fig.~\ref{csectionD} that the production of charginos and neutralinos is
the dominant source of sparticles at the LHC. Gluinos and squarks are
relatively heavy so that their production becomes rapidly suppressed as
$\Lambda$ is increased. Slepton and sneutrino production forms only a
percent of the total SUSY cross section.  As already noted, the
branching fraction for the two body decay $\tz_1 \to \tG h$ exceeds 50\%
if $\Lambda \agt 80$~TeV, and becomes $\sim$75\% for $\Lambda \agt
200$~TeV \cite{tevatron}.  The decays $\tz_1 \to Z\tG$ account for
essentially all the remaining decays of $\tz_1$ since $\tz_1 \to
\gamma\tG$ with a branching fraction of at most a few if $\Lambda \geq
120$~TeV. The heavier neutralino $\tz_2$ as well as the lighter chargino
$\tw_1$ decay via three body decays, 
but their decay products are soft since they are only a little heavier
than $\tz_1$. As a result, 
signatures for $\tw_1\tw_1$ or $\tz_i\tw_1$ production closely resemble
those for $\tz_1\tz_1$ production.

We first consider the reach via the canonical missing $E_T$~\cite{lhc1}
and multilepton analyses~\cite{lhc2}. The results of our computation are
shown in Fig.~\ref{reachmlD}. This figure corresponds to
Fig.~\ref{reachmlB} for model line B and, in fact, is qualitatively
quite similar to it.  Despite the fact that the $\tz_1$ decays mainly to the
light Higgs boson for $\Lambda \ge 120$~TeV, the best reach, which is
obtained in the $n_{\ell} \geq 3$ leptons channel, extends out to
$\Lambda = 140$~TeV corresponding to $m_{\tg} = 2$~TeV, with squarks
only slightly heavier. The reach in 1 lepton and 2 OS leptons channels
are comparable, up to $\Lambda = 130$~TeV, while the zero lepton channel
again gives the poorest reach. To understand this, we first note that
although for these large values of $\Lambda$ chargino and neutralino
production dominates the total cross section, gluino and squark
production events (which can be detected much more efficiently) form a
significant portion of the cross section after the selection cuts:
events from the lighter chargino and $\tz_{1,2}$ production (these have
masses of just $\sim 300$~GeV) pass the
hard selection cuts with very low efficiency since $m_{\tw_1} \sim
m_{\tz_1} \sim m_{\tz_2} \sim 300$~GeV.  Gluinos decay to
$t\bar{t}\tz_{1,2}$ or $tb\tw_1$ with a branching fraction of 60\%,
with the bulk of the remaining decays into the heavier chargino and to the
heaviest neutralino.  Squarks mainly decay to the gaugino-like $\tw_2,
\tz_3$ and $\tz_4$ ($\tq_R$ mainly decays to $\tz_3$); $\tw_2$ and
$\tz_4$ mainly decay via two body decays to vector bosons or the
lightest Higgs scalar $h$ and a lighter chargino or neutralino, and at a
few percent into left sleptons and sneutrinos. The
$\tz_3$, on the other hand, mostly decays via $\tz_3 \to \tell_R\ell$.
For $\Lambda=140$~GeV, $\tell_L$ and $\tnu$ mainly cascade
decay to $\tz_3$ which then mostly decays into right sleptons.
Notice that if $\tell_L$, $\tnu$ or $\tz_3$ are produced at any step,
their decays are by themselves likely to give two hard leptons,  resulting
in multiple leptons from squark and gluino events. For the higgsino NLSP
case, the cascade decay chains are more complicated than in the
canonical mSUGRA case, because of the dominance of gluino and squark
decays to the heavier charginos and neutralinos, and in the gluino case,
also to top quarks.

For values of $\Lambda$ close to the LHC reach in Fig.~\ref{reachmlD},
the NLSP dominantly decays via $\tz_1 \to \tG h$. In this case, since
$h$ mainly decays via $h \to b\bar{b}$, the SUSY signal will contain
$b$-jets in addition to leptons, $\eslt$ and possibly photons (if one of
the NLSPs decays via the photon mode which has a branching fraction of a
few percent \cite{tevatron}). We were, therefore, led to examine whether
requiring tagged $b$-jets in SUSY events can increase the reach via the
observation of lighter chargino and neutralino events which have a
production cross section $\geq 300$~$fb$.  The dominant SM background to
multi-$b$ plus large $\eslt$ 
events presumably comes from $t\bar{t}$ production. We
examined several event topologies as well as distributions from the
signal and from the top background and concluded that the most promising
channel was the one with just two tagged $b$-jets channel, with the following
cuts to reduce the top background:
\begin {enumerate}
\item $\eslt \ge 100$~GeV;
\item $p_T(b-jet) > 50$~GeV;
\item $\eslt + \Sigma E_{Tj} \ge 1500$~GeV, where the sum extends over
untagged jets.
\end{enumerate}

The results of our computation are shown in Table~II. 
We show the signal cross sections 
for $\Lambda =140, 150$ and 160~TeV and compare these with the top
background for two choices for the $b$ mistagging probability. The set
of columns labelled 2\% assume the $b$ mis-tagging probability
introduced in Sec.~II. The columns labelled 1\% assume the mis-tagging
probability to be half this. The statistical significance $N_S
\over \sqrt{N_B}$ is for 10~$fb^{-1}$. We see from the Table that for the
canonical mis-tag rate even the $\Lambda=150$~TeV case falls just short of
observability but appears to pass our observability criteria with the
more optimistic assumptions about $b$-mistagging. In view of the fact
that $t\bar{t}$ is the only background we have included 
in our computation, and further, that the reach is only marginally
increased above this background, we conclude that the
canonical multilepton search is sufficient to probe model line D at the
LHC, and further that the LHC reach extends out to $\Lambda=140-145$~TeV
(corresponding to $m_{\tg} =2$~TeV, with squarks a little heavier). 

We have also examined the $\geq 3$ tagged $b$-jet channel which without the
hardness cut 3 has a much better $S/B$ ratio. The signal cross
section, in this channel, is small and it is difficult to make
other cuts (to reduce the top background) without reducing the signal to
below the level of observability. It is conceivable though that with an
integrated luminosity of 100~$fb^{-1}$ this channel may prove more
promising; the background from $4b$ production would then have to be
carefully examined.

Finally, we examine whether it is possible to identify the Higgs boson,
which is the characteristic feature of model line D, in SUSY events. It
has been known for many years that the presence of Higgs bosons can
be inferred either via the multiplicity of $b$-jets in SUSY events
\cite{bbtw}, or more directly, via the reconstruction of a mass bump in
the $M_{bb}$ distribution~\cite{lhc1,frank}. We follow the latter
approach, and attempt to reconstruct the lighter Higgs boson (which has
a mass of about 104~GeV) in SUSY events in channels with identified
$b$-jets for $\Lambda = 80$~TeV and $\Lambda = 110$~TeV. Again top quark
pair production is the major SM background. To reduce this, we require
that 
\begin{displaymath}
\eslt + \Sigma E_T(jets) \geq E_0,
\end{displaymath}
where the sum extends over untagged $b$-jets with $E_T \geq 100$~GeV.
We adjust the cut variable $E_0$ depending on $\Lambda$ and
on the event topology. In particular, we choose $E_0=1300$ (900)~GeV for
two (three) tagged $b$ events if $\Lambda=80$~TeV, and 1500 (1100)~GeV
if $\Lambda=110$~TeV. We found that we were unable to obtain a mass bump
(with these as well as several other selection cuts that we examined)
even though the cross section was quite large. We traced this to the
fact that most of the time one of the two $b$-jets from $h$ decay
usually has $p_T \leq 50$~GeV, and so fails our tagging criteria. In
other words, the two $b$-jets from the decay of $h$ are almost never
simultaneously tagged. To enable the identification of $h$ in SUSY
events, we include in our analysis that $b$-jets with $E_T$ between 25
and 50~GeV may also be tagged, but with a reduced efficiency which we
take to be~\cite{ratio} $0.015\times p_T-0.25$. We extrapolate our previous
mistagging rate down to 25~GeV.

The $M_{bb}$ distribution in the two and three tagged $b$ channels after
the cuts above is shown in Fig.~\ref{mbb_mlD} for $\Lambda=80$ and
110~TeV. The dashed histograms are the $t\bar{t}$ background while the
solid histograms are the sum of the signal and background. We see that
there is a clear peak close to $m_h$ in the $M_{bb}$ distributions in
the two tagged $b$ channel. The distribution cuts off at the low end
because each tagged jet has at least 25~GeV, while the continuum beyond
the peak corresponds to combinations where one or both $b$-jets come
from sources other than Higgs decay or are QCD jets that have been
mistagged.  The corresponding peaks for the three $b$ channel, though
present, is not as distinctive, presumably because the combinatorial
background also contributes to the continuum.

We also examined the distribution of $p_T$ of the Higgs boson
reconstructed from two tagged $b$s with $M_{bb} \simeq m_h$; {\it i.e.} in
the peak. We might expect that this distribution would scale with
$m_{\tz_1}$, and hence provide a measure of $\Lambda$. We found,
however, that the distributions for $\Lambda$ in the 80-110~TeV range
overlapped quite substantially, and conclude that it is difficult to
measure $\Lambda$ this way. We attribute this to the fact that
the change in $m_{\tz_1}$ is rather limited (160-220~GeV), and further,
that $h$ can also be produced at an earlier stage in the cascade,
thereby smearing this distribution. For
larger values of $\Lambda$ the event rate appears to be too low for this
to be feasible.

\subsection{Model Line E: Higgsino NLSP with $\tz_1\to Z\tG$}

If the NLSP contains a significant fraction of the SUSY partner of the
would-be neutral Goldstone boson (that becomes the longitudinal
component of the $Z$ by the Higgs mechanism), it would preferentially
decay via $\tz_1 \to Z\tG$ provided that this decay is not kinematically
suppressed. It has recently been pointed out that this situation is
indeed realized~\cite{mt} for a Higgsino model line where the sign of
$\mu$ is flipped relative to that in model line D above. We thus
consider a final model line with $n_5=2$, $\tan\beta=3$, $M/\Lambda =3$,
$C_{grav}=1$ but fix $\mu =\frac{3}{4}M_1$. The decay pattern of the
neutralino NLSP is shown in Fig.~\ref{NLSPEdecay}. We see that for
$\Lambda \geq 85$~TeV (corresponding to an NLSP heavier than 140~GeV)
the decay $\tz_1 \to Z\tG$ dominates, while for smaller masses, the
photon decay is the main mode.~\footnote{In this case the search
strategy would be similar to that for model line A, or via the inclusive
$Z\gamma$ search discussed at the end of this section.} The branching fraction for the decay
$\tz_1\to h\tG$ becomes significant only when $\Lambda$ is rather large.

The main sparticle production cross sections are shown in
Fig.~\ref{csectionE} which is qualitatively very similar to
Fig.~\ref{csectionD}. This should not be surprising because changing the
sign of $\mu$ qualitatively affects only the NLSP decay pattern, though
of course there is some effect on the chargino and neutralino masses and
mixing angles. Again, as in model line D, the mass gap between $\tw_1$
or $\tz_2$ and $\tz_1$ is not large, so that $\tw_1$ and $\tz_2$ decays
are similar to those from the decay of just the NLSP because the
remaining decay products tend to be soft.

The reach of the LHC via the canonical $\eslt$ and multilepton channels
is illustrated in Fig.~\ref{reachmlE}. As before, the number on the
curves denotes the $E_T^c$ value that optimizes the signal. Again, the
upper (lower) solid curve denotes the trilepton (zero lepton) plus
$\eslt$ channel. The main difference from Fig.~\ref{reachmlD} is that
the greatest reach, which again extends out to 140~TeV (corresponding to
$m_{\tg}=2$~TeV, with squarks marginally heavier), is now obtained via
the single lepton channel. It may seem surprising that the
reach via multilepton signals degrades relative to model line D where
NLSP decays are not expected to lead to isolated leptons. To understand
this, we examined the decay patterns of sparticles (for
$\Lambda=140$~TeV), and found that, except for $\tz_1$, these are very
similar for these two model lines. The big difference is that the
branching fraction for $\tz_3 \to \tell_R\ell$ which was 26\% per lepton
flavour for model line D is now about 19\%, while there is a
corresponding increase in $\tz_3 \to \tw_1 W$ decay. Since the presence
of $\tz_3$ in a cascade decay chain frequently results in two leptons
(see the corresponding discussion for model line D), it is not
surprising that the inclusive trilepton signal is reduced relative
to model line D. Indeed we have checked that for $E_T^c \geq 200$~GeV,
this signal, although it has a large statistical significance, falls
below the five event level (for $\Lambda \geq 140$~TeV), so that we are
forced to lower $E_T^c$ to 100~GeV for which the SM
background~\cite{lhc2} is much larger. \footnote{We mention that with
20~$fb^{-1}$ the trilepton channel would again yield the best reach.}

We now examine whether the LHC reach can be extended by using the
feature that NLSP mostly decays via $\tz_1 \to \tG Z$, so that a
significant fraction of SUSY events would contain a real $Z$ boson in
addition to jets, leptons and $\eslt$ \cite{wood}. Since the $Z$ can be
identified via its leptonic decay, we focus on events with an identified
$Z$ together with either two jets with $E_{Tj} \geq 50$~GeV
or additional leptons. To further suppress backgrounds, we found it was
necessary to require a large $\eslt$ in these events. The signal cross
sections for three values of $\Lambda$ just beyond the reach in
Fig.~\ref{reachmlE}, together with the main 
sources of SM backgrounds, are shown in Table~III for $\eslt \ge
230$~GeV. Here $\ell^+\ell^-$ pairs with $|M(\ell^+\ell^-)-M_Z| \leq
10$~GeV are identified as a $Z$. We see from this Table that the
dominant background comes from $t\bar{t}$ production where the leptons
from top decay accidentally reconstruct the $Z$ mass. We also see that with
these cuts the signal falls below the observable level, though it should
be noted that all these cases would be observable for an integrated
luminosity $\sim 35-40$~$fb^{-1}$. Two comments are worth noting.
\begin{itemize}
\item The signal cross section for two identified $Z$ bosons falls below
the observable level. This can even be seen from Fig.~\ref{csectionE}
and Fig.~\ref{NLSPEdecay}
where the total SUSY production cross section ($\sim 330$~$fb$) and
branching ratio (75\%) for $\tz_1 \to Z\tG$ for
$\Lambda=150$~TeV yields in $330 \times (0.75)^2 \times (0.06)^2 \times 10
\simeq 7$~events even before any acceptance cuts.

\item The dominant background from tops in Table~III mainly contributes
to the $Z$ plus two jet channel, and is completely negligible
for $Z$ plus lepton channels. We have checked though that 
the $Z$ plus two jet channel accounts for  about 70-80\% of the
signal, so that the total signal in just the
$Z$ plus lepton channels again falls below the observable level.

\end{itemize}

Motivated by the fact that the branching fraction for $\tz_1 \to \gamma
\tG$ is $\sim$10\% for $\Lambda \sim 140$~TeV, we have also examined
whether inclusive $Z\gamma$ signals could extend the LHC reach. Toward
this end we computed SM backgrounds from $Zj$, $WZ+ZZ$ and $t\bar{t}$
production, where a photon from radiation or the decay of a hadron is
accidently isolated, using ISAJET. We also estimated the background from
$Z\gamma$ production (this process is not included in ISAJET) by
introducing the $2 \to 2$ $q\bar{q} \to Z\gamma$ process into a private
version of ISAJET, and ignoring spin correlations for leptons from $Z$
decay. We search for events in the inclusive $Z\gamma$ channel where the
$Z$ is identified by its leptonic decay and the isolated photon has
$E_{T\gamma} \geq 25$~GeV.  We also require $\eslt \geq 100$~GeV. This
last cut completely eliminates the physics background from $Z\gamma$ events
(which essentially cuts off by $\eslt \simeq 40$~GeV) and reduces
backgrounds from other sources. The results of our calculation are
summarized in Table~IV. We see that the physics backgrounds that we have
computed are completely negligible.

We expect though that the dominant background could well come from
instrumental effects when a jet or lepton is misidentified as a photon.
Although a calculation of this detector-dependent background is beyond
the scope of the present analysis, we attempt to obtain a rough estimate
by computing the inclusive $Z$ background for $\eslt \geq 100$~GeV and
then multiplying this by the probability for a jet to fake a photon. We
found this inclusive $Z$ background cross section to be 192~$fb$
from $t\bar{t}$ production, 36.7~$fb$ from $Zj$ production and 10.2~$fb$
from $WZ+ZZ$ production, giving a total of 239~$fb$. Assuming that the
probability for a jet to fake a photon is $\sim 2 \times
10^{-4}$~\cite{private}, and further that there are typically $\sim 5$
(10) jets with $E_T \geq 25$~GeV in these LHC events, we obtain the
``fake $\gamma + Z$'' background cross section of $239 \times 2\times
10^{-4} \times 5(10) \sim 0.24 (0.47)$~$fb$. We see from Table~IV that
this reducible background can potentially be comparable to the
signal. Furthermore, since this estimate does not include the
effects of cracks and other real-world detector effects, it is likely to
be increased once these effects are incorporated. We conclude that {\it
unless these reducible backgrounds can be controlled} it is unlikely
that the inclusive $Z\gamma$ channel will increase the reach. But with
excellent jet-$\gamma$ rejection, the LHC reach for model line E could
be as large as $\sim 155$~TeV. Moreover, if the signal is truly
rate-limited (which does not seem to be the case with our assumptions
about the instrumental backgrounds), the reach would grow significantly
with the size of the data sample.

\section{Summary and Concluding Remarks}

The CERN LHC is generally considered as the accelerator
facility at which weak scale supersymmetry will either be discovered or
definitely excluded. This is because the LHC experiments are
expected~\cite{lhc1,lhc2,ATLAS,CMS,ratio} to probe $m_{\tg}$ out to 2~TeV, at
least within the mSUGRA framework. This extends well beyond the
generally accepted bounds from naturalness considerations. Moreover,
experiments at the LHC should be able~\cite{rpv} to probe gluinos and
squarks beyond 1~TeV even in the experimentally unfavourable scenario
where $R$-parity violating interactions cause the neutralino LSP to
decay hadronically, thereby reducing the $\eslt$ as well as degrading
the lepton isolation in SUSY events. Although not in themselves a
``proof'', it is precisely such studies that have led to the general
belief that the LHC will be the definitive machine that will test SUSY.
In this study, we have added to this evidence by examining the SUSY
reach of the LHC within the GMSB framework with SUSY broken at a scale
of a few hundred TeV. In this case, the gravitino is then the LSP and
SUSY signatures are governed by the nature of the NLSP which then decays
into the gravitino and SM particle(s). In our analysis, we
conservatively assume that the NLSP decay is prompt, and do not attempt
to use the possible presence of a displaced vertex to enhance the SUSY
signal. 

We divide our study along various model lines introduced in
Section~I. For each of these model lines, the NLSP (which is almost
always produced as
the penultimate product of the sparticle decay cascade) decays via
characteristic modes, which can then be used to enhance the SUSY signal,
or even to narrow down the model parameters. We do not attempt the
latter in the present study. For model line A which was studied in
Ref.~\cite{bmtw}, for instance, $\tz_1 \to \gamma\tG$, so that the
presence of hard, isolated photons in SUSY events served to 
distinguish signal from SM
backgrounds, resulting in a SUSY reach that corresponding to $m_{\tg}
\leq 2.8$~TeV, to be compared with a reach of about 2~TeV expected
within the mSUGRA framework. 

In this paper we have examined the LHC reach for four other model lines
corresponding to (B) the stau NLSP scenario, (C) the co-NLSP scenario
where all three flavours of light charged sleptons are essentially
degenerate, and two higgsino-like NLSP scenarios where NLSP decays to
(D) the light Higgs boson and (E) the $Z$ boson, dominate. The usual
strategy for SUSY searches via $\eslt$ and multilepton events will probe
gluino masses up to about 2~TeV, except in the co-NLSP model line where
the fact that selectrons and smuons are produced in many cascade decay
chains leads to a plethora of $\eslt$ events with $\geq 4$ isolated
leptons ($e$ or $\mu$) which have very small SM backgrounds. In this
case the reach of the LHC extends to $m_{\tg} = 3$~TeV!  For model lines B and
D, however, we found that whereas the decays of the NLSP may result in
events that are characteristic of the particular model line, signals
with identified $\tau$ leptons or tagged $b$ jets do not, in
general, extend the SUSY reach beyond what is obtained via standard
multilepton analyses. For model line E, some enhancement of the reach
might be possible via
the inclusive $Z\gamma$ channel which is rate-limited,
but only if superb jet-$\gamma$ discrimination is possible.
Nevertheless, it should be kept in mind that it will be worthwhile
to search for these characteristic events in any LHC data sample enriched in
SUSY events, as they will point to the underlying theory, and possibly
also lead to a measurement of some of the model parameters~\cite{frank2}.

\acknowledgements
This research was supported in part by the U.~S. Department of Energy
grants DE-FG02-97ER41022 and DE-FG03-94ER40833. During his stay at the
University of Hawaii where this work was begun, P.M. was partially
supported by Funda\c{c}\~ao de Amparo \`a Pesquisa do Estado de
S\~ao Paulo (FAPESP).

%
%


\begin{table}
\caption[]{The $\Lambda = 120$~TeV and $\Lambda = 160$~TeV signal cross
section in $fb$ for multiple 
$\tau$-jets plus lepton plus $\eslt$ events for model line B.}
 
\label{bback}
\bigskip
\begin{tabular}{|c|cccc|cccc|}
 & \multicolumn{4}{c |} {$\Lambda=120$~TeV } &\multicolumn{4}{c |}{$\Lambda= 160$~TeV } \cr
%
 &  $0 \tau$ & $ 1 \tau$ & $2 \tau$ & $\geq 3 \tau$ & $0 \tau$ & $ 1 \tau$ & $2 \tau$ & $ \geq 3 \tau$   \cr
\tableline
$0\ell$ &  1.75 & 1.55 & 0.62 & 0.02 & 0.16 & 0.21 & 0.09 & 0.01 \cr
$1\ell$ & 5.55 & 4.37 & 1.24 & 0.16 & 0.69 & 0.75 & 0.26 & 0.04  \cr
$2 OS\ell$ & 2.09 & 1.47 & 0.27 & 0.00 & 0.30 & 0.29 & 0.09 & 0.02 \cr
$2 SS\ell$ & 1.56 & 1.19 & 0.39 & 0.01 & 0.31 & 0.23 & 0.09 & 0.01 \cr
$\geq3\ell$ & 4.23 & 2.08 & 0.34 & 0.00 & 0.75 & 0.51 & 0.13 & 0.01 \cr
%
\end{tabular}
\end{table}

 

\begin{table}
\caption[]{ The signal and $t\bar{t}$ background cross sections in $fb$
and the statistical significance $N_S \over \sqrt{N_B}$, assuming an
integrated luminosity of 10~$fb^{-1}$, via the 2 b channel for model
line D. We show the results for two choices of mistagging rates as
discussed in the text.}
 
\label{jetty}
\bigskip
\begin{tabular}{|c|ccc|ccc|}
$\Lambda(TeV)$ &  $\sigma_S(2\%)$ &
$\sigma_B(2\%)$ & $N_S \over \sqrt{N_B}$(2\%) &  $\sigma_S(1\%)$ &
$\sigma_B(1\%)$ & $N_S \over \sqrt{N_B}$(1\%) \cr
\tableline
140   & 2.11 & 0.89 & 7.09  & 2.05 & 0.72 & 7.62\cr
\hline
150   & 1.34 & 0.89 &4.49 & 1.36 & 0.72 & 5.36 \cr
\hline
160   & 0.70 & 0.89 & 2.34  & 0.73 & 0.72 & 2.72 \cr
\end{tabular}
\end{table}

\begin{table}
\caption[]{The SUSY signal cross sections (in $fb$) for inclusive $Z$
plus $\eslt \geq 230$~GeV events for model line E together with SM
backgrounds from $Zj$, $WZ+ZZ$ and $t\bar{t}$ production. For the case of
the $t\bar{t}$ background, the `$Z$' is a fake from two leptons
accidently reconstructing the $Z$ mass. The signal cross
sections are shown in the first three columns for $\Lambda =140, 150$
and 160~TeV along with the statistical significance for an integrated
luminosity of 10~$fb^{-1}$.}
 
\label{ZmlE}
\bigskip
\begin{tabular}{|c|ccc|ccc|}
& $\Lambda(140)$ & $\Lambda(150)$ & $\Lambda(160)$ & $Zj$ & $WZ+ZZ$ &
$t\bar{t}$ \\
\tableline
$\sigma (fb)$ & 2.73 & 2.50 & 2.22 & 2.45 & 0.84 & 4.13 \\
\hline
$N_S/\sqrt{N_B}$ & 3.17 & 2.90 & 2.58 & & & 

\end{tabular}
\end{table}

\begin{table}
\caption[]{The SUSY signal cross sections (in $fb$) for the inclusive
$Z\gamma$ channel for model line E together with physics SM backgrounds
from $Zj$, $WZ+ZZ$, $t\bar{t}$ and $Z\gamma$ production. Here an
$\ell^+\ell^-$ pair with a mass within 10~GeV of $M_Z$ is identified as
a $Z$. The signal cross sections are shown for $\Lambda=140, 150$ and
160~TeV. See also the discussion in the text about potential
instrumental backgrounds to this signal. }
 
\label{zgamma}
\bigskip
\begin{tabular}{|c|ccc|cccc|}
& $\Lambda(140)$ & $\Lambda(150)$ & $\Lambda(160)$ & $Zj$ & $WZ+ZZ$ &
$t\bar{t}$ & $Z\gamma$\\
\tableline
$\sigma (fb)$ & 0.78 & 0.59 & 0.40 & 0.0048 & 0.0031 & 0 & 1.5$\times10^{-4}$

\end{tabular}
\end{table}

%
    
\iftightenlines\else\newpage\fi
\iftightenlines\global\firstfigfalse\fi
\def\dofig#1#2{\iftightenlines\epsfxsize=#1\centerline{\epsfbox{#2}}\bigskip\fi}

\begin{figure}
\dofig{5.3in}{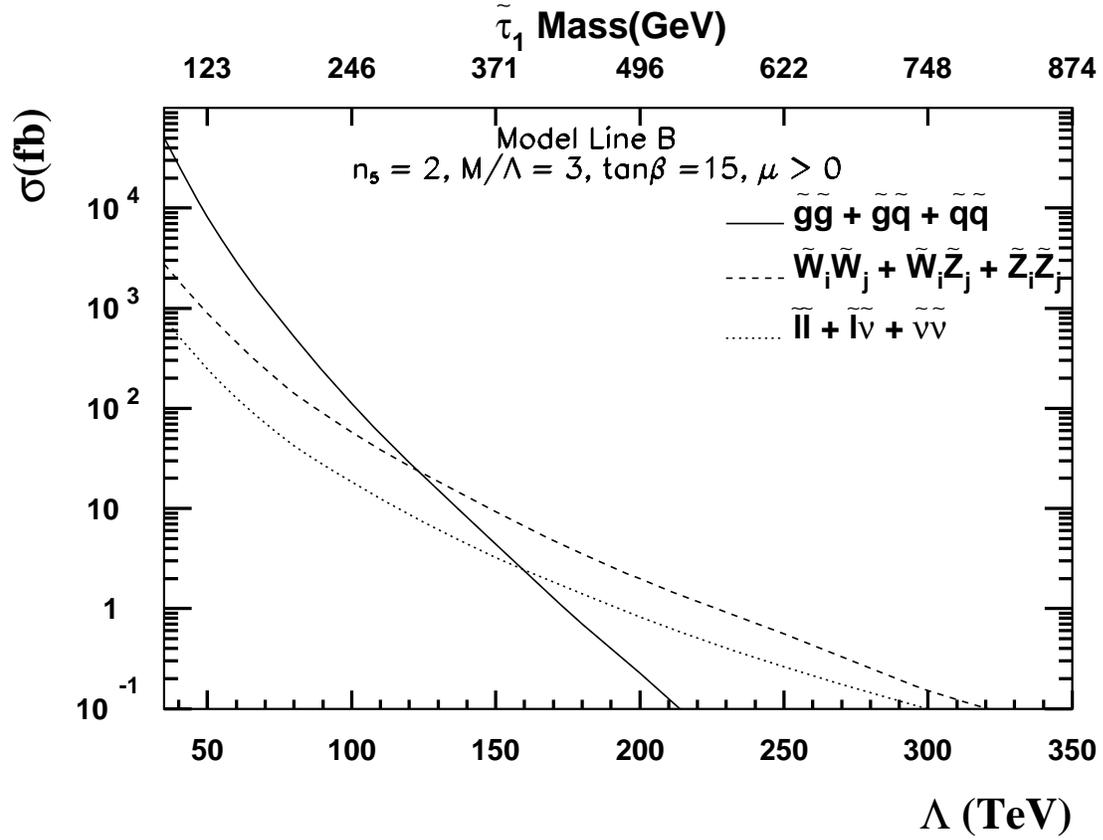}
\caption[]{Sparticle production cross sections versus $\Lambda$ for various
sets of SUSY processes at a 14~TeV $pp$ collider within the GMSB
framework for the $\ttau_1$ NLSP model line B.
For sleptons, we sum over the first two families.
}
\label{csectionB}
\end{figure}

\begin{figure}
\dofig{6in}{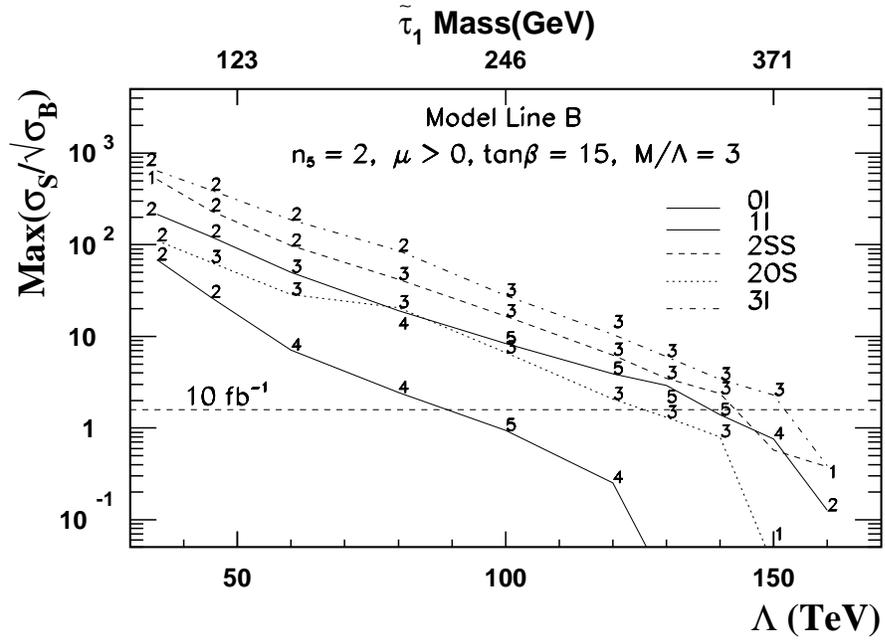}
\caption[]{ We show the $\max\ \sigma_S/\sqrt{\sigma_B}$ with respect to 
$E^c_T$ versus $\Lambda$ for model line B for $ 0 \ell+\eslt$ and the
various multilepton channels introduced in the text. We evaluate
$\sigma_S/\sqrt{\sigma_B}$ for $E_T^c=100, 200, 300, 400$ and 500~GeV
and choose the value of $E_T^c$ that maximizes this, subject to a
minimum signal cross section of 0.5~$fb$. For each value of $\Lambda$
for which we evaluate the signal, the number 
denotes this optimal choice of
$E^c_T$, {\it e.g}  2 denotes $E^c_T = 200$~GeV, etc. The horizontal
line shows the minimum cross section for a $5\sigma$ signal, assuming an
integrated luminosity of 10~$fb^{-1}$. }
\label{reachmlB}
\end{figure}

\begin{figure}
\dofig{6in}{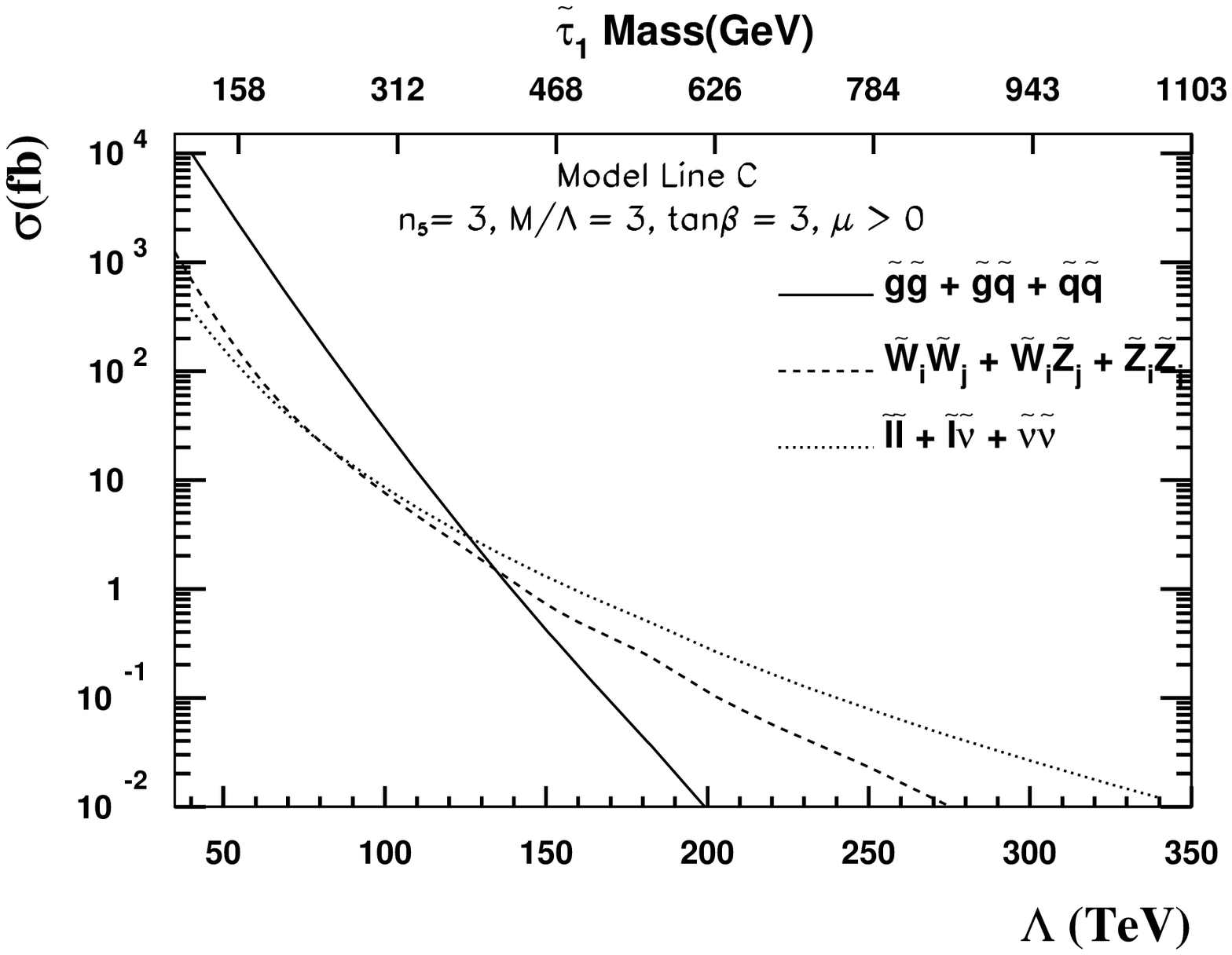}
\caption[]{ The same as Fig.~\ref{csectionB} except that the cross
sections are now shown for model line C.}
\label{csectionC}
\end{figure}

\begin{figure}
\dofig{7in}{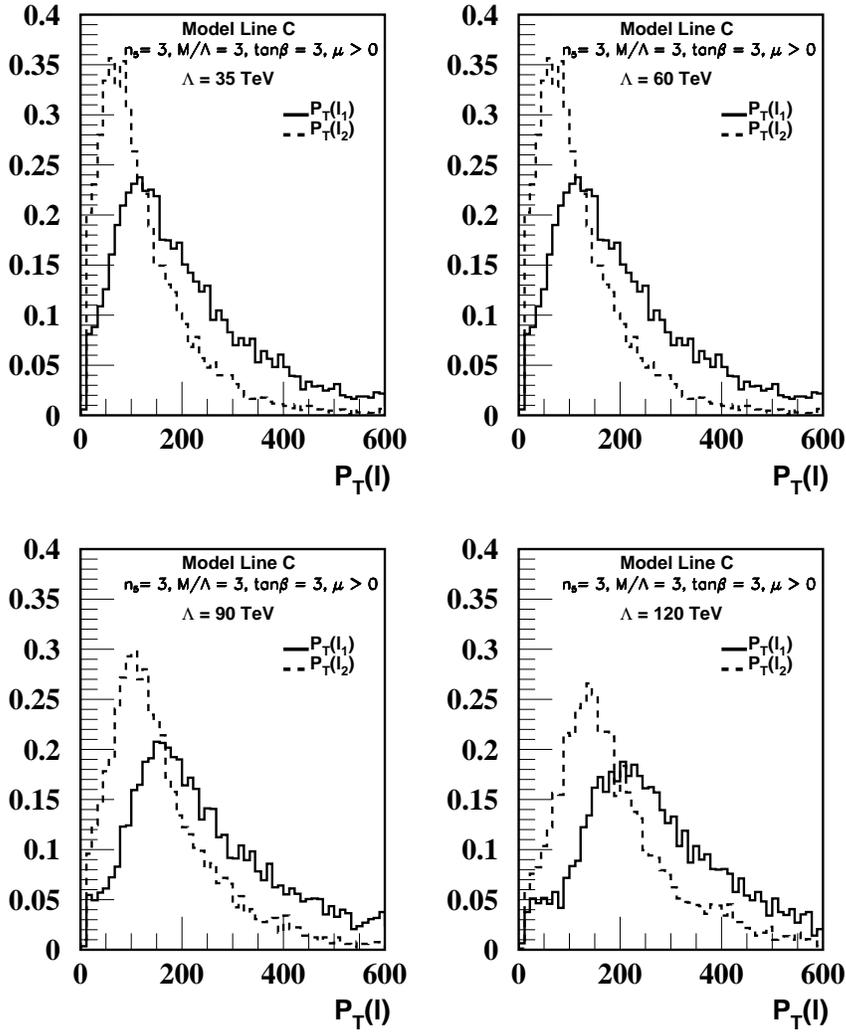}
\caption[]{The transverse momentum distributions for the hardest (solid)
and second hardest (dashed) leptons for four choices of $\Lambda$ for
model line C.}
\label{mlC_lep12pt}
\end{figure}

\begin{figure}
\dofig{5in}{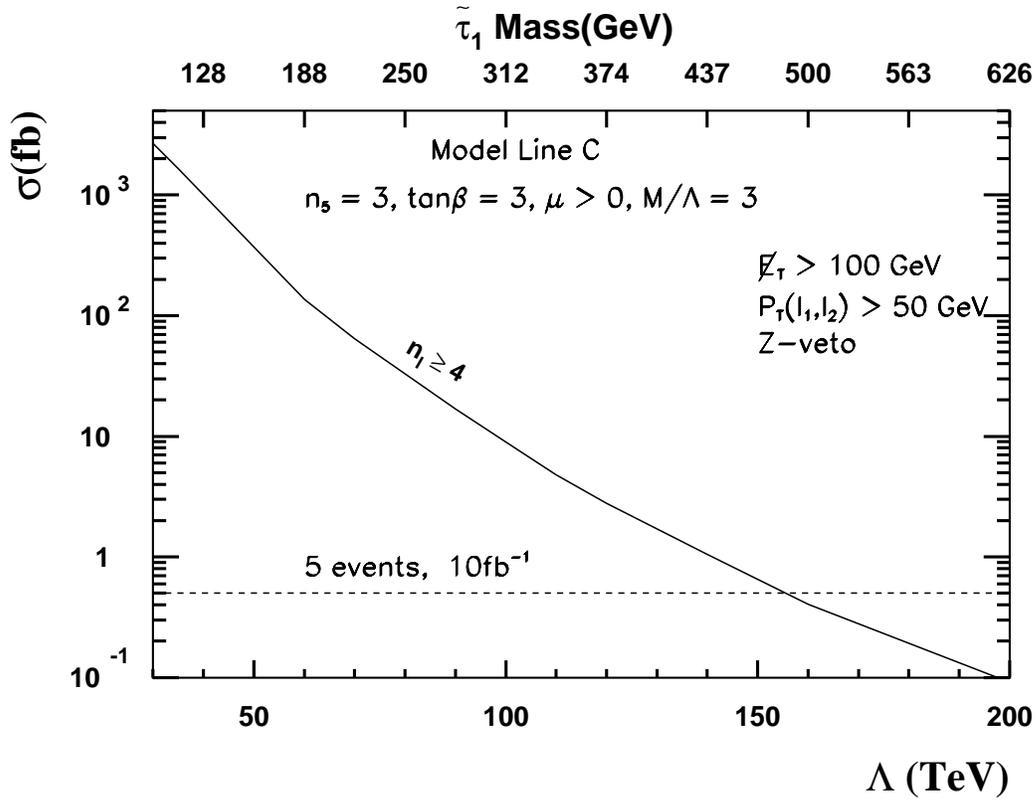}
\caption[]{The SUSY signal cross section for the inclusive $n_l
\geq 4$ channel for model line C after all the cuts discussed in
the text. The SM background in this channel is negligible.
The horizontal line denotes the
cross section corresponding to the 5 event level for an integrated
luminosity of 10~$fb^{-1}$.}
\label{reachmlC}
\end{figure}

\begin{figure}
\dofig{6in}{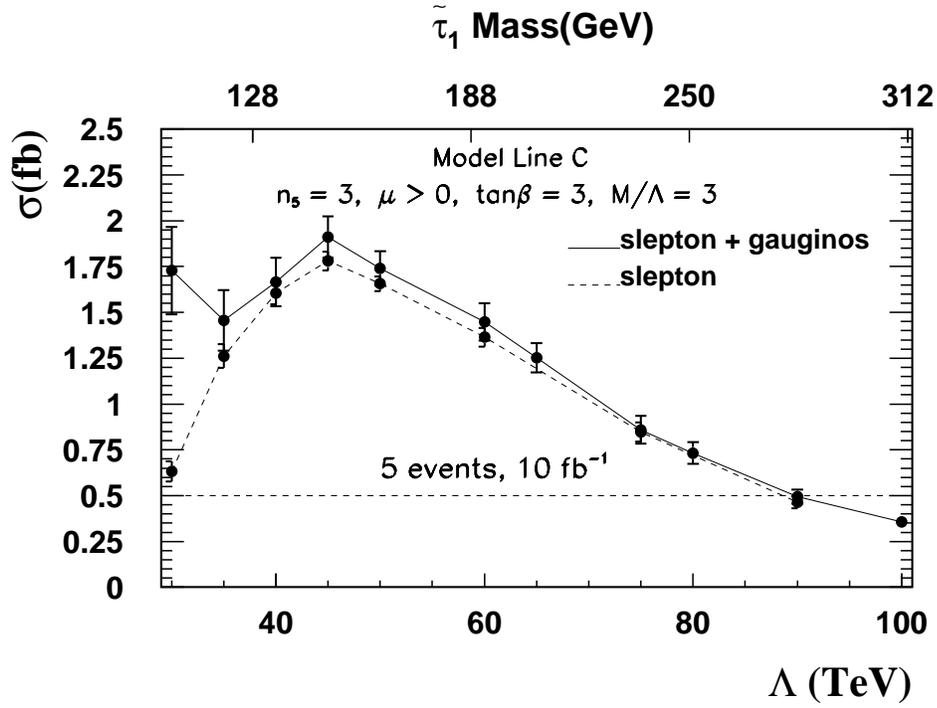}
\caption[]{The signal cross section from slepton pair production
(dashed) and slepton, chargino and neutralino production (solid) in the
opposite sign dilepton channel after cuts designed to optimize the
slepton signal for model line
C. The  dots correspond to the values of $\Lambda$ for which we did the
simulation. The horizontal line denotes the minimum observable level
for a $5\sigma$ signal at the LHC assuming an integrated luminosity of
10~$fb^{-1}$. The reach of $\Lambda=90$~TeV corresponds to
$m(\tell_R)=280$~GeV. }
\label{reachslepC}
\end{figure}

\begin{figure}
\dofig{6in}{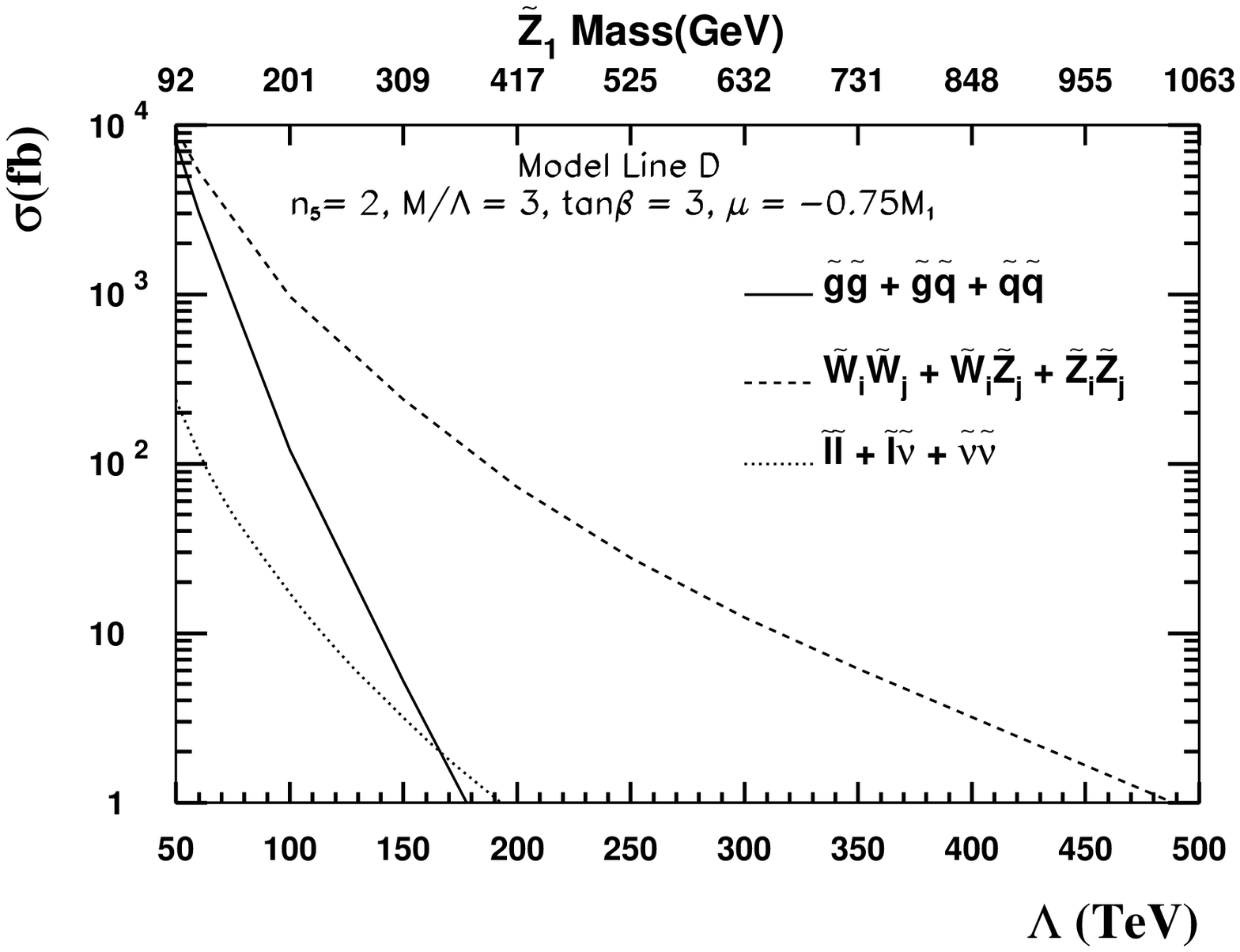}
\caption[]{ The same as Fig.~\ref{csectionB} except that the cross
sections are now shown for model line D.}
\label{csectionD}
\end{figure}

\begin{figure}
\dofig{7in}{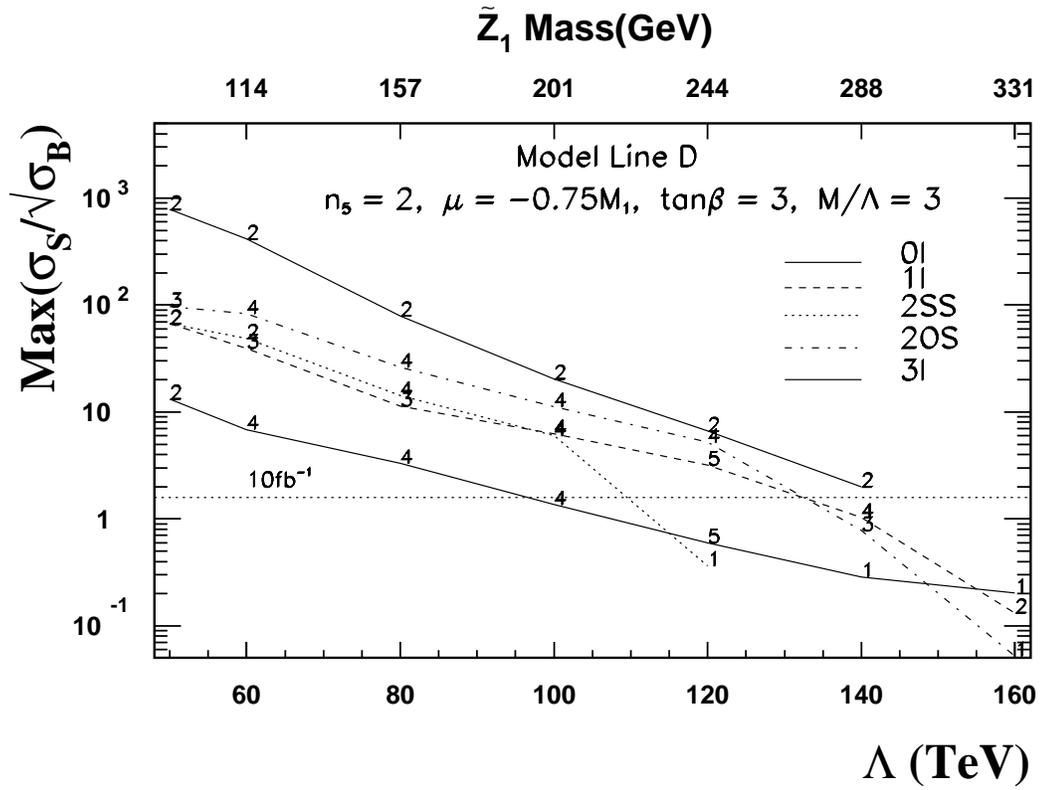}
\caption[]{The same as Fig. \ref{reachmlB}, but for the higgsino NLSP
model line D. Notice though that the legends for the various channels
are different.}
\label{reachmlD}
\end{figure}

\begin{figure}
\dofig{7in}{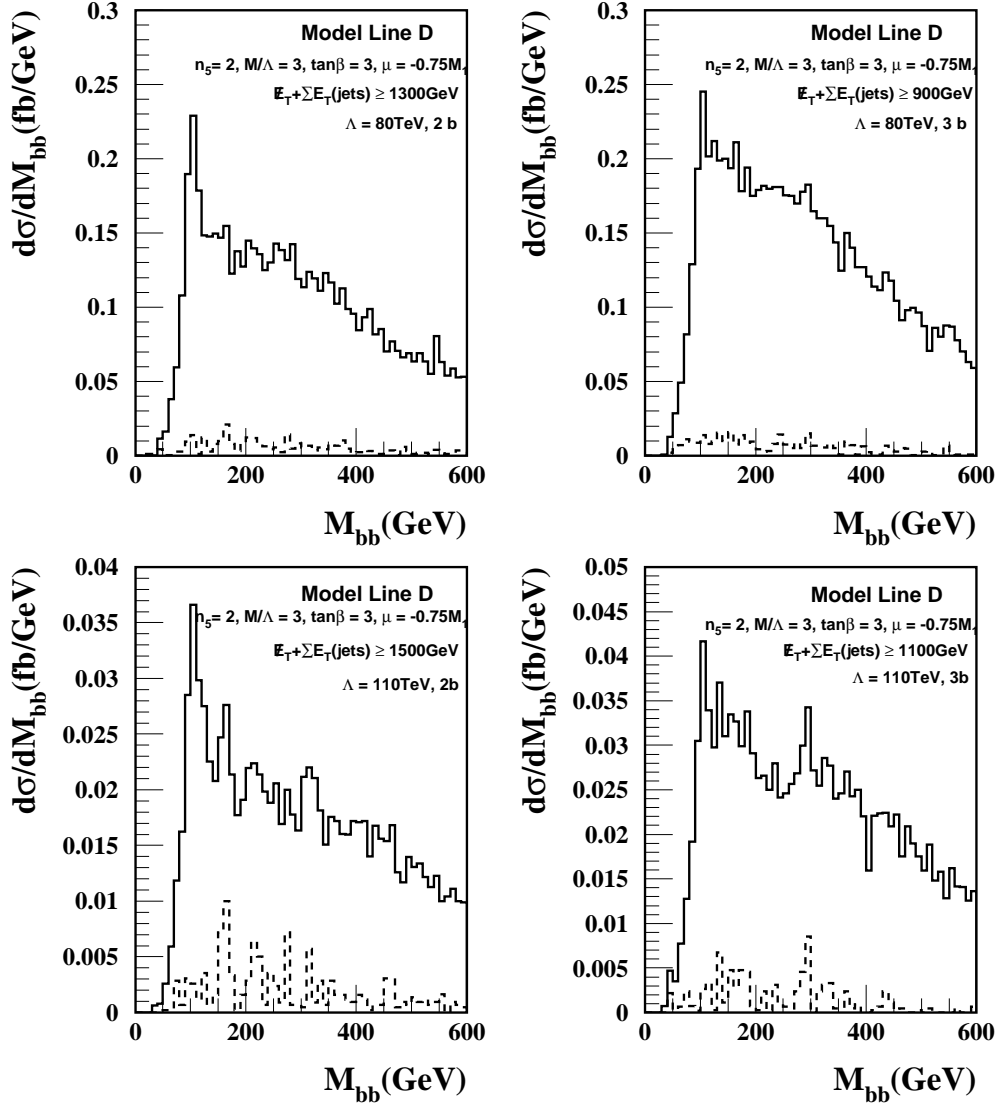}
\caption[]{The $M_{bb}$ distribution for events after all cuts discussed
in the text in the two tagged $b$ and 3 tagged $b$ channels for the
higgsino model line D. We illustrate the results for $\Lambda=80$ and
110~TeV. The solid histogram is the signal plus background, while the
dashed line is just the $t\bar{t}$ background.  }
\label{mbb_mlD}
\end{figure}

\begin{figure}
\dofig{7in}{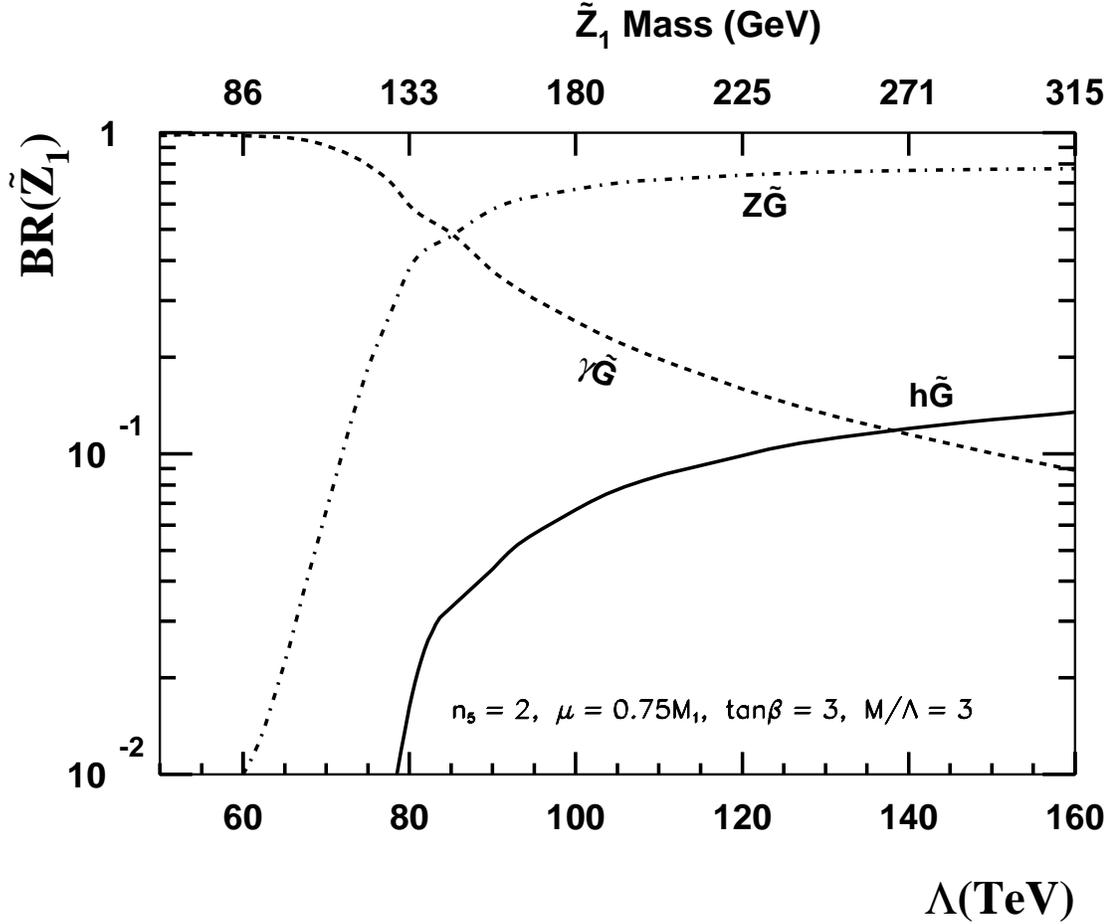}
\caption[]{The branching fractions for various decays of the neutralino NLSP in
model line E versus the parameter $\Lambda$. The scale on top shows the
NLSP mass.}
\label{NLSPEdecay}
\end{figure}

\begin{figure}
\dofig{6in}{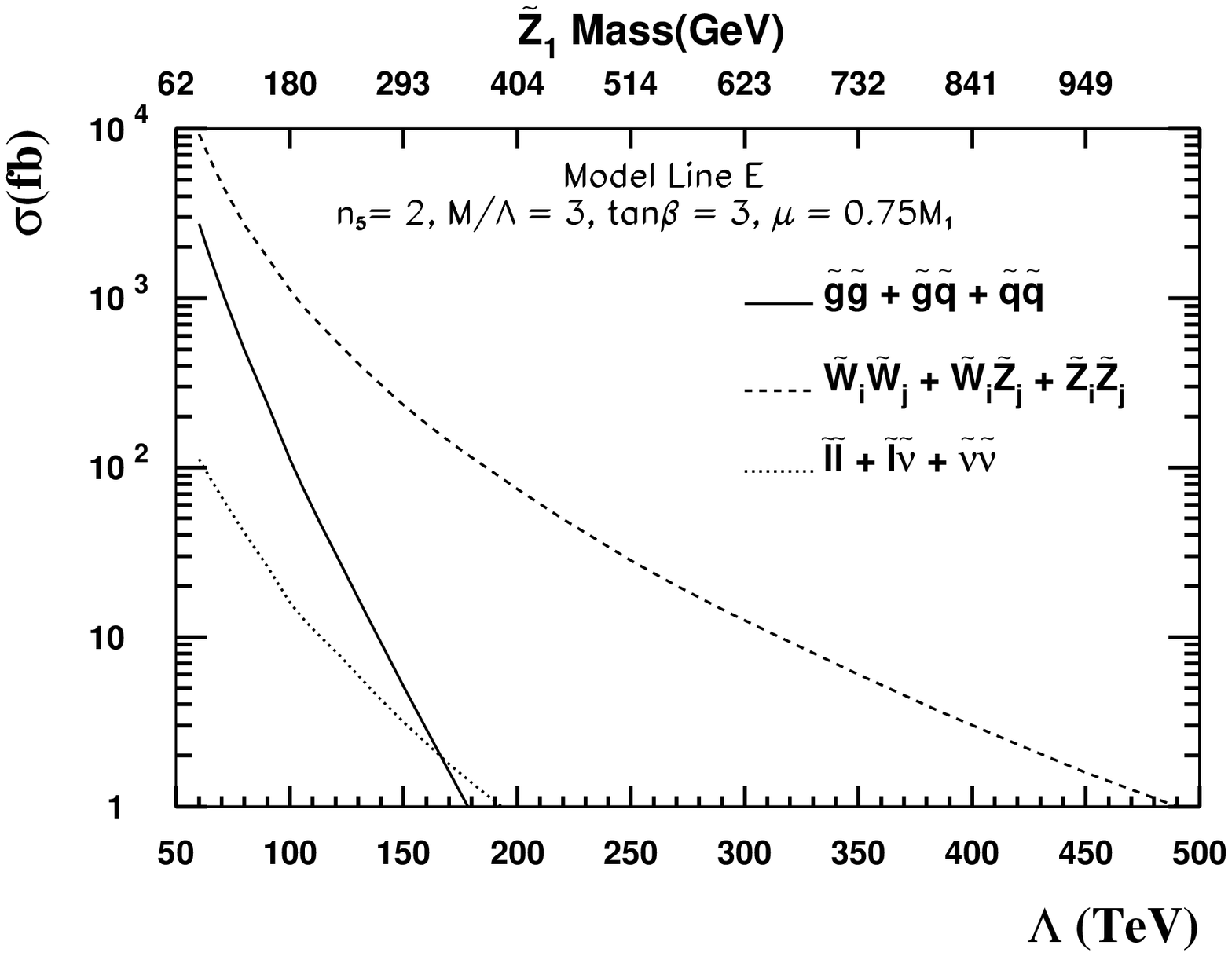}
\caption[]{The same as Fig.~\ref{csectionB} except that the cross
sections are now shown for model line E.} 
\label{csectionE}
\end{figure}

\begin{figure}
\dofig{7in}{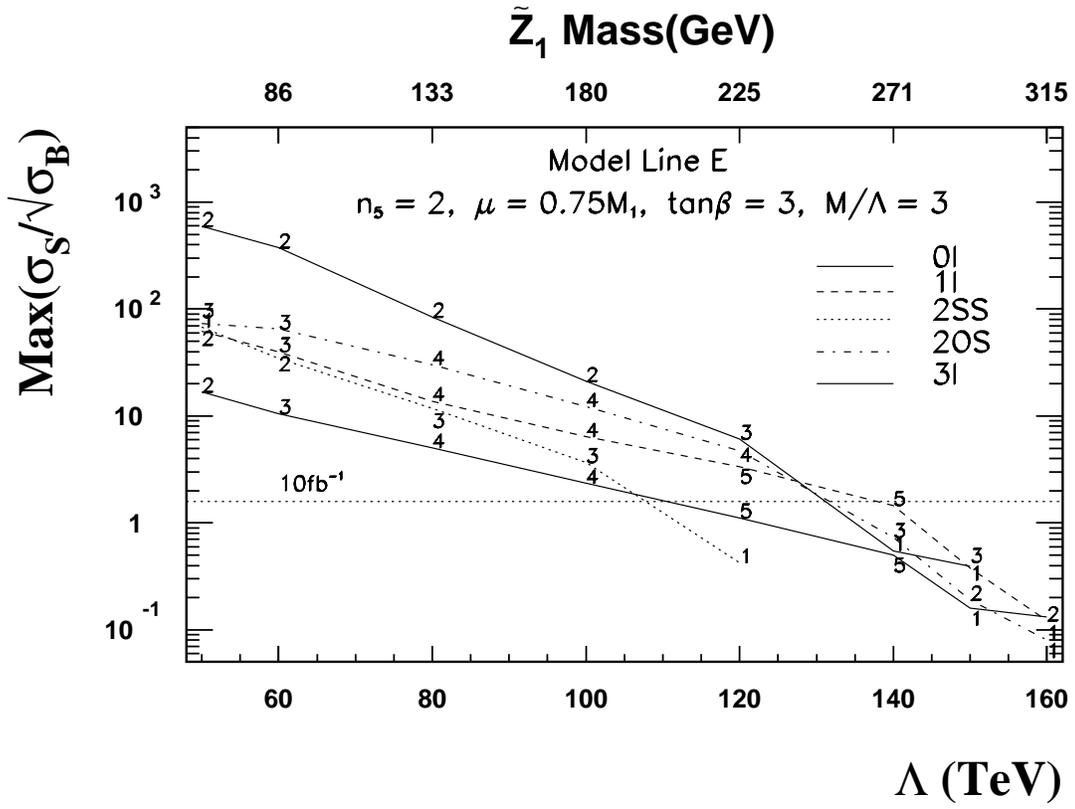}
\caption[]{The same as Fig.~\ref{reachmlD}, but for the higgsino model
line E.}
\label{reachmlE}
\end{figure}

\end{document}